\documentclass[aps,pra,superscriptaddress,amsfonts,amsmath,amssymb,reprint,showpacs,floatfix]{revtex4-1}

\usepackage{url}
\usepackage{bm}
\usepackage{graphicx}
\usepackage{amsmath}
\usepackage{amstext}
\usepackage{amssymb}
\usepackage{amsfonts}
\usepackage{amsbsy}
\usepackage{verbatim}
\usepackage{color}
\usepackage[colorlinks=true, urlcolor=blue, linkcolor=blue, citecolor=blue ]{hyperref}
\usepackage{multirow}

\begin{document}

\title{Spiral antiferromagnets beyond the spin-wave approximation: frustrated $XY$ and Heisenberg models
in the honeycomb lattice}

\author{Andrea Di Ciolo}
\affiliation{Joint Quantum Institute and Department of Physics, University of Maryland, College Park, 20742, USA}
\author{Juan Carrasquilla}
\affiliation{Perimeter Institute for Theoretical Physics, Waterloo, Ontario, N2L 2Y5, Canada}
\affiliation{Department of Physics, The Pennsylvania State University, University Park, Pennsylvania 16802, USA}
\author{Federico Becca}
\affiliation{Democritos National Simulation Center, Istituto Officina dei Materiali del CNR and SISSA-International 
School for Advanced Studies, Via Bonomea 265, I-34136 Trieste, Italy}
\author{Marcos Rigol}
\affiliation{Department of Physics, The Pennsylvania State University, University Park, Pennsylvania 16802, USA}
\author{Victor Galitski}
\affiliation{Joint Quantum Institute and Department of Physics, University of Maryland, College Park, 20742, USA}

\begin{abstract}
We examine the stability of classical states with a generic incommensurate spiral order against quantum 
fluctuations. Specifically, we focus on the frustrated spin-1/2 $XY$ and Heisenberg models on the 
honeycomb lattice with nearest-neighbor $J_1$ and next-nearest-neighbor $J_2$ antiferromagnetic couplings. 
Our variational approach is based on the Jastrow wave functions, which include quantum correlations on top 
of classical spin waves. We perform a systematic optimization of wave vectors and Jastrow pseudo-potentials 
within this class of variational states and find that quantum fluctuations favor collinear states over 
generic coplanar spirals. The N\'eel state with ${\bf Q}=(0,0)$ extends its stability well beyond the 
classical value $J_2/J_1=1/6$. Most importantly, the collinear states with ${\bf Q}=(0,2\pi/\sqrt{3})$ 
(and the two symmetry-related states) are found to be stable in a large regime with intermediate frustration, 
while at the classical level they are limited to the point $J_2/J_1=0.5$. For large frustration, the 
$120^\circ$ state is stabilized for finite values of $J_2/J_1$ in both models. 
\end{abstract}

%\date{\today}


\maketitle

\section{Introduction}\label{sect:intro}

Magnetic frustration in spin systems is responsible for complex phase diagrams due to the 
competition between states that are very close in energy but exhibit fundamentally different 
properties~\cite{BookMila}. Although exotic phases with no magnetic order (possibly having 
topological order and fractional excitations) represent the forefront of present research in the 
field, understanding complex magnetically ordered phases in quantum spin models remains of great 
interest as well. Indeed, at present spin liquids are found in only relatively small regions of 
frustrated spin models, such as the $J_1$-$J_2$ model on the square lattice~\cite{Jiang2012,Hu2013} 
or the Heisenberg model on the kagome lattice~\cite{Yan2011,Iqbal2013}. On the other hand, ordered 
phases are ubiquitous and represent important examples of correlated states~\cite{BookMila}.

The first step and simplest approximation in describing these phases is obtained by considering spins 
as classical variables, thus completely neglecting quantum fluctuations. This approximation is adequate 
when the spin $S$ is large, e.g., for half-filled $d$ or $f$ shells in the presence of a large Hund
coupling (mathematically speaking, this approximation becomes exact when $S=\infty$).
In order to include quantum corrections, a systematic perturbative approach can be constructed by using 
the so-called Holstein-Primakoff transformation~\cite{Holstein1940}. Here, the first-order terms at the 
$O(1/S)$ level already contain quantum correlations that correctly describe the low-energy spin-wave 
spectrum. In addition, within this scheme, it is possible to obtain rather accurate results for the 
renormalization of the magnetization due to quantum fluctuations. The $O(1/S)$ quantum corrections
may also select the correct ground state when the classical ground state is highly degenerate,
e.g., for the $J_1$-$J_2$ model on the square lattice for $J_2/J_1>0.5$~\cite{Chandra1990}. 
Unfortunately, this technique becomes very cumbersome when considering higher corrections beyond 
$O(1/S)$. Alternative approaches exist, e.g., a modified spin-wave theory~\cite{Takahashi1989}, 
but they too involve complicated perturbative expansions. Therefore, identification of simple variational 
wave functions is useful to go beyond the spin-wave approximation and capture non-perturbative effects. 
Indeed, this approach has been widely used to study different spin models on frustrated 
lattices~\cite{Liang1988,Huse1988,Manousakis1991,Boninsegni1996,Franjic1997}. 

Among frustrated spin models, a particularly interesting example is the $J_1$-$J_2$ Heisenberg model 
on the honeycomb lattice:
\begin{equation}\label{eq:HHeis}
H_{\rm Heis} = J_1 \sum_{\langle ij \rangle} {\bf S}_i \cdot {\bf S}_j 
             + J_2 \sum_{\langle\langle ij \rangle\rangle} {\bf S}_i \cdot {\bf S}_j,
\end{equation}
where the sums $\langle ij \rangle$ and $\langle\langle ij \rangle\rangle$ run over all the 
nearest-neighbor and next-nearest-neighbor bonds, respectively; ${\bf S}_i = (S_i^x,S_i^y,S_i^z)$ 
is the spin operator on site $i$. In the following, we consider ${\bf a}=(1,0)$ and 
${\bf b}=(-1/2,\sqrt{3}/2)$ as the two Bravais vectors defining the honeycomb lattice. 

In recent years, a lot of effort has focused on understanding the phase diagram of this model, 
mainly in the case of antiferromagnetic Heisenberg interactions~\cite{Mulder2010,Clark2011,
Albuquerque2011,Farnell2011,Reuther2011,Oitmaa2011,Mezzacapo2012,Ganesh2013,Zhu2013a,Gong2013}. 
Interestingly, the $J_1$-$J_2$ model on the honeycomb lattice has a non-trivial phase diagram even 
at the classical level, since for $J_2/J_1>1/6$ the ground state has an infinite degeneracy due to 
the fact that the wave vector ${\bf Q}$ can vary on closed contours in the Brillouin zone; see
below~\cite{Rastelli1979,Fouet2001}. The spin-1/2 quantum model has been the subject of recent works,
and was shown to exhibit magnetically disordered regions for sufficiently large frustration, which 
presumably have plaquette and dimer orders~\cite{Ganesh2013,Zhu2013a,Gong2013}.

A closely related model that also exhibits a rich phase diagram is the frustrated spin-1/2 $XY$ model
in the honeycomb geometry:
\begin{equation}\label{eq:Hxy}
H_{\scriptscriptstyle XY}=J_1\!\sum_{\langle ij \rangle}(S_i^x S_j^x + S_i^y S_j^y  )
       + J_2\!\sum_{\langle\langle ij \rangle\rangle} (S_i^x S_j^x + S_i^y S_j^y).
\end{equation}
By using the fact that $S_j^{\pm} = S_j^x \pm iS_j^y$, the spin-1/2 $XY$ model is equivalent to
``non-interacting'' hard-core bosons with hopping amplitudes $J_1/2$ and $J_2/2$ at nearest- and
next-nearest-neighbor sites, respectively~\cite{Varney2011,Varney2012}. By using exact
diagonalizations on small clusters, it was suggested that a disordered spin-liquid phase may
appear in a narrow regime of intermediate frustration, in between magnetically ordered antiferromagnetic 
and collinear phases~\cite{Varney2011}. This scenario is consistent with variational calculations 
involving partonic wave functions~\cite{Carrasquilla2013}. However, recent density-matrix renormalization
group calculations have instead pointed toward an unexpected ordered phase (with spin order along 
the $z$-axis, which is equivalent to a charge-density wave in the boson language)~\cite{Zhu2013b}.
Both these scenarios are highly unusual and currently the nature of the ground state for 
$0.2 \lesssim J_2/J_1 \lesssim 0.3$ is not understood.

Both exact-diagonalization and density-matrix renormalization group studies are done in finite 
systems with highly constrained geometries. The complex quantum states that these methods 
select are in competition with other ordered states, including spirals that are often suggested by the 
classical analysis. The stabilization of these spirals may be highly frustrated in finite systems, 
which generates concerns about extrapolating finite-size results to the thermodynamic limit. Therefore, 
it is important to carry out a systematic study of these spiral states on large lattices and including 
quantum fluctuation around classical ground states. This is the focus of our work.

Our approach is based on introducing Jastrow wave functions that are particularly suitable to describe 
magnetically ordered states~\cite{Manousakis1991,Franjic1997}. These variational states are 
constructed by applying a long-range Jastrow factor, which enables us to account for quantum effects,
to classical spin waves with a given wave vector ${\bf Q}$ and relative phase $\eta$ between the
two spins in the unit cell (the detailed description is given in Sec.~\ref{sec:wavefunction}). 
We show that the best Jastrow state may have a ${\bf Q}$ vector different from the one that 
minimizes the classical energy. In general, collinear phases are highly favored over generic spiral 
ones. In particular, the N\'eel state with ${\bf Q}=(0,0)$ remains stable up to $J_2/J_1 \simeq 0.3$
in the Heisenberg model and $J_2/J_1 \simeq 0.26$ for the $XY$ model (to be compared with 
$J_2/J_1 = 1/6$ for the classical model). Moreover, a collinear state with ${\bf Q}=(0,2\pi/\sqrt{3})$
(and the other two symmetry-related wave vectors) is remarkably stable in a wide region of the phase 
diagram, namely, $0.7 \lesssim J_2/J_1 \lesssim 1.4$ for the Heisenberg model and 
$0.26 \lesssim J_2/J_1 \lesssim 1$ for the $XY$ model. While incommensurate spiral states are clearly 
defeated for small and intermediate values of $J_2/J_1$, they may survive in a relatively small range 
of frustration, before the $120^\circ$ state sets in.

The paper is organized as follows: In Sec.~\ref{sec:classical}, we provide a short summary of
the classical results. In Sec.~\ref{sec:wavefunction}, we discuss the form of the variational
states that are used. In Sec.~\ref{sec:results}, we show our numerical results, and, finally,
in Sec.~\ref{sec:conc}, we draw our conclusions.

\section{Classical results}\label{sec:classical}

Here, we briefly summarize the classical results~\cite{Rastelli1979,Mulder2010}, which apply for
both Heisenberg and $XY$ models. Assuming coplanar order in the $XY$ plane, the spins on the two 
sublattices are
\begin{equation}
{\bf S}_i=S \Big [ \cos({\bf Q} \cdot {\bf R}_i),\sin({\bf Q} \cdot {\bf R}_i),0 \Big ]
\end{equation}
when the site $i$ belongs to the $\mathcal{A}$ sublattice, and
\begin{equation}
{\bf S}_i=-S \Big [ \cos({\bf Q} \cdot {\bf R}_i+\eta),\sin({\bf Q} \cdot {\bf R}_i+\eta),0 \Big ]
\end{equation}
when the site $i$ belongs to the $\mathcal{B}$ sublattice. Here, ${\bf R}_i$ denotes the coordinates
of the site $i$ in the triangular Bravais lattice, the two sites in the unit cell having the same 
${\bf R}_i$. ${\bf Q}$ is the spiral wave vector, and $\eta+\pi$ (notice the definition with the minus 
sign for spins on the $\mathcal{B}$ sublattice) defines the angle between the two spins on different 
sublattices. Within this notation, the N\'eel antiferromagnet is described by ${\bf Q}={\bf \Gamma}=(0,0)$
and $\eta=0$ while the state with $120^\circ$ order has ${\bf Q}={\bf K}=(4\pi/3,0)$ or 
${\bf K'}=(2\pi/3,2\pi/\sqrt{3})$ and an arbitrary phase shift (the two sublattices being totally 
decoupled).
%%%%%%%%%%%%%%%%%%%%%%%%%%%%%%
\begin{figure}[!b]
\includegraphics[width=1.0\columnwidth]{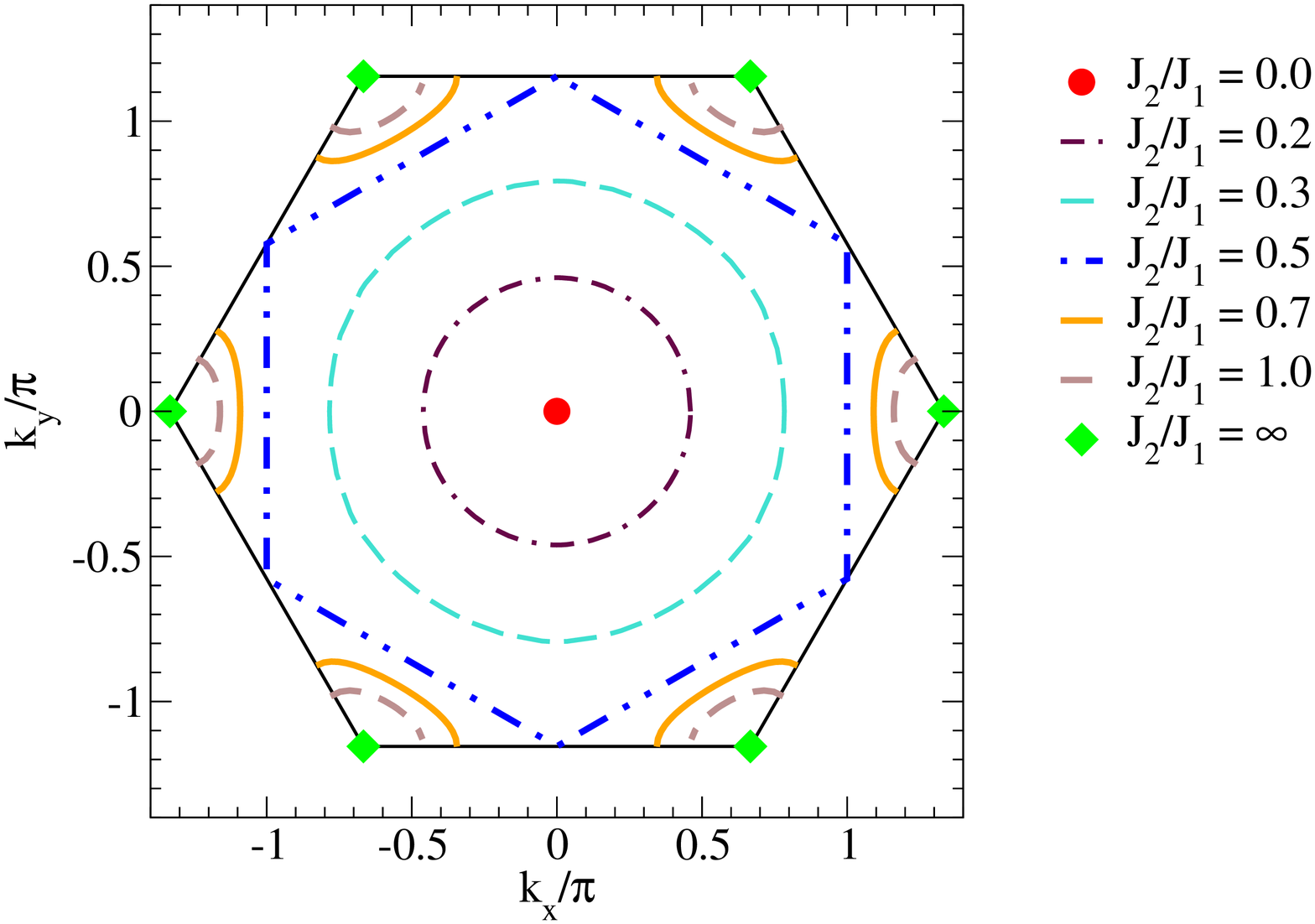} %QClas.eps}
\caption{\label{fig:QClass}
(Color online) The manifold of classically degenerate spiral wave vectors for a few values of
$J_2/J_1$. For $J_2/J_1<1/6$, the lowest-energy state has ${\bf Q}^*=(0,0)$, while for
$J_2/J_1>1/6$ the ground state is degenerate: the ${\bf Q}^*$ wave vectors form closed
contours around the ${\bf \Gamma}$ point for $1/6<J_2/J_1<1/2$ and around the ${\bf K}$ (or ${\bf K'}$)
point for $J_2/J_1>1/2$. For $J_2/J_1 \to \infty$, ${\bf Q}^*={\bf K}$ or ${\bf K'}$.}
\end{figure}
%%%%%%%%%%%%%%%%%%%%%%%%%%%%%%
The classical energy per spin for a generic coplanar spin wave can be written as
\begin{eqnarray}
E_{\rm cl}&=&-\frac{J_1S^2}{2} \Big [ \cos\eta + \cos(\eta-Q_b) + \cos(\eta-Q_a-Q_b) \Big] \nonumber \\ 
          &&+J_2S^2 \Big[\cos Q_a + \cos Q_b + \cos (Q_a+Q_b) \Big], 
\label{Ecl}
\end{eqnarray}
where $Q_a={\bf Q} \cdot {\bf a}=Q_x$ and $Q_b={\bf Q} \cdot {\bf b}=-Q_x/2+\sqrt{3}Q_y/2$.
By minimizing the classical energy with respect to $Q_a$, $Q_b$, and $\eta$, one finds two
regimes: for $J_2/J_1<1/6$, the lowest-energy state has ${\bf Q}^*=(0,0)$ and $\eta^*=0$, while 
for $J_2/J_1>1/6$ it has a finite ${\bf Q}^*$ satisfying the relation
\begin{equation}\label{eq:Qclass}
\cos Q_a^* + \cos Q_b^* + \cos(Q_a^*+Q_b^*) = \frac{1}{2} \left[ \left(\frac{J_1}{2J_2}\right)^2-3 \right],
\end{equation}
while $\eta^*$ is completely defined by
\begin{eqnarray}
&&\sin \eta^* = \frac{2J_2}{J_1} \left[ \sin Q_b^* + \sin(Q_a^*+Q_b^*) \right], \\
&&\cos \eta^* = \frac{2J_2}{J_1} \left[ 1+ \cos Q_b^* + \cos(Q_a^*+Q_b^*) \right].
\end{eqnarray}
Given Eq.~\eqref{eq:Qclass}, there are infinite spiral wave vectors that minimize the energy at 
any given value of $J_2/J_1>1/6$.

\begin{figure}[!h]
\includegraphics[width=1.0\columnwidth]{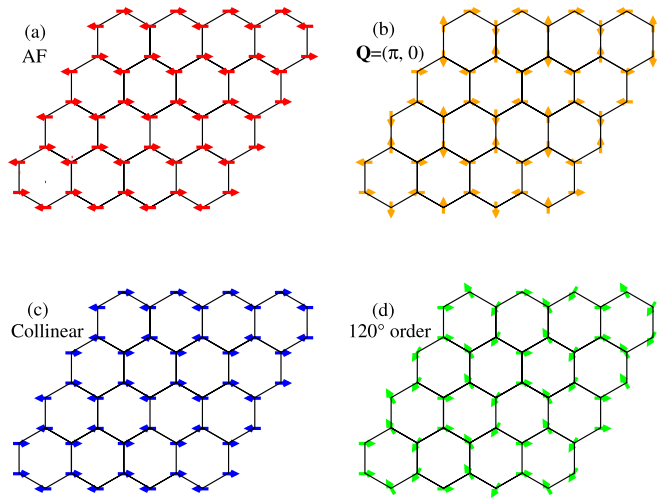}
\caption{\label{fig:realspins}
(Color online) Real-space illustration of some of the ordered states considered in this work.
(a) N\'eel antiferromagnet, for which ${\bf Q}={\bf \Gamma}=(0,0)$ and $\eta=0$.
(b) $(\pi,0)$-spiral, for which ${\bf Q}=(\pi,0)$ and $\eta=0$. 
(c) Collinear state, for which ${\bf Q}={\bf M}=(0,2\pi/\sqrt{3})$ and $\eta=\pi$.
(d) $120^\circ$ ordered phase, for ${\bf Q}={\bf K}$ or ${\bf K'}$ and $\eta$ arbitrary.}
\end{figure}

In Fig.~\ref{fig:QClass}, we report the classically degenerate solutions for a few values of $J_2/J_1$: 
they form closed contours around ${\bf \Gamma}$ for $1/6 <J_2/J_1<1/2$ (spirals I) and around 
${\bf K}$ or ${\bf K'}$ for $J_2/J_1>1/2$ (spirals II). For $J_2/J_1=1/2$, the closed contour has a 
hexagonal shape and, among all possible spirals, there are three particularly simple collinear states, 
in which all nearest-neighbor bonds along one direction are ferromagnetic, while the other two are 
antiferromagnetic. One of these states has ${\bf Q}={\bf M}=(0,2\pi/\sqrt{3})$ and $\eta=\pi$. These 
collinear states are stabilized in a wide region of the phase diagram when quantum fluctuations are 
considered. The $120^\circ$ state is recovered only when $J_2/J_1 \to \infty$, i.e., when the two 
sublattices are totally decoupled. Real-space spin configurations for some representative states described 
above are depicted in Fig.~\ref{fig:realspins}. 

Finally, we mention the fact that the $O(1/S)$ quantum corrections lift the huge classical degeneracy
for $J_2/J_1>1/6$, selecting wave vectors along ${\bf \Gamma}$-${\bf M}$ (and symmetry-related ones)
for $J_2/J_1<1/2$ and along the border zone for $J_2/J_1>1/2$~\cite{Mulder2010}. In the following, 
we will analyze the extent to which the $O(1/S)$ scenario is preserved for $S=1/2$ models.

%%%%%%%%%%%%%%%%%%%%%%%%%%%%%%
%%%%%%%%%%%%%%%%%%%%%%%%%%%%%% 

\section{Variational wave functions}\label{sec:wavefunction} 

The variational states containing quantum fluctuations are defined by: 
\begin{equation}\label{eq:Psiansatz}
|\Psi\rangle = \mathcal{J}_z \mathcal{P}_{S^z_{\rm tot}=0} |{\rm SW}\rangle.
\end{equation}
Here, $|{\rm SW}\rangle$ is a spin-wave state, described by a wave vector ${\bf Q}$ and a 
phase shift $\eta$:
\begin{eqnarray}
|{\rm SW}\rangle 
&=& \prod_{i} \left(|\downarrow\rangle_i + e^{\imath ({\bf Q} \cdot {\bf R}_i + \eta_i)}
|\uparrow\rangle_i \right) \nonumber \\
&=& \prod_{i} e^{\imath ({\bf Q} \cdot {\bf R}_i + \eta_i) (S^z_i+1/2)} 
\left(|\downarrow\rangle_i + |\uparrow\rangle_i \right),
\end{eqnarray}
where $\eta_i=0$ if $i$ belongs to sublattice $\mathcal{A}$ and $\eta_i=\eta+\pi$ if it 
belongs to sublattice $\mathcal{B}$. $|{\rm SW}\rangle$ is equivalent to a classical state where each spin
points in a given direction in the $XY$ plane. $\mathcal{P}_{S^z_{\rm tot}=0}$ is the projector onto 
the subspace with $S^z=0$. When projected into this subspace, $|{\rm SW}\rangle$ is translationally 
invariant, since for example:
\begin{equation}
{\mathcal T}_{\bf a} |{\rm SW}\rangle = e^{\imath {\bf Q} \cdot {\bf a} (S^z_{\rm tot} +N)} 
|{\rm SW}\rangle,
\end{equation}
where ${\mathcal T}_{\bf a}$ is the translational operator of one lattice site along ${\bf a}$ and 
$N$ is the number of sites of the Bravais lattice, ${\bf Q} \cdot {\bf a} N =0$ (mod $2\pi$).

Quantum fluctuations are included through the long-range Jastrow factor
\begin{equation}
\mathcal{J}_z= \exp\left(\frac{1}{2} \sum_{ij} v_{ij}S^z_i S^z_j \right),
\end{equation}
where, in a translationally invariant system, the pseudopotential $v_{ij}$ depends upon the vector 
${\bf R}_i-{\bf R}_j$. All the independent parameters (i.e., the $v_{ij}$'s as well 
as ${\bf Q}$ and $\eta$) are optimized via Monte Carlo simulations in order to minimize the variational
energy~\cite{Sorella2005,Yunoki2006}.

Quantum fluctuations in the spin-1/2 case are expected to be strong enough to substantially change the 
classical scenario. We would like to emphasize that we consider only two-body correlations; higher-order 
terms have been used to improve both the signs and the amplitudes of variational states in frustrated 
lattices~\cite{Huse1988}. Indeed, while our wave function has the correct signs for the unfrustrated 
case with $J_2=0$ (where a Marshall sign rule holds), in general it does not reproduce the correct 
(and unknown) sign structure. We would like also to remark that the variational state explicitly 
breaks the spin SU$(2)$ symmetry of the Heisenberg model. This is apparent from the fact that the order 
parameter is in the $XY$ plane, and the Jastrow factor contains only the $z$ component of the spin operator. 

Since quantum effects may favor states with a different ordering vector than the one selected 
by the classical model, we compute the energy by optimizing the Jastrow parameters for all 
non-equivalent spin waves allowed by the particular lattice size and study an extensive range of 
values of $\eta$ in order to determine the state with the lowest possible energy. It should be 
emphasized that not all possible ${\bf Q}$ vectors are accessible on finite clusters.
This fact is particularly relevant when considering spirals, which may be frustrated on a given 
finite lattice; nonetheless, it is still possible to follow the evolution of the wave vector as a 
function of the frustrating ratio $J_2/J_1$ and to detect the stability of collinear phases.

Finally, it is important to note that classical states with generic ${\bf Q}$ do not in general 
possess the underlying rotational invariance of the lattice, i.e., pairs of spins along different 
spatial orientations and the same geometric distance apart might be correlated in different ways. When 
attempting to build correlations on top of such states, those differences are naturally accounted for 
in our trial states by considering a Jastrow factor with parameters $v_{ij}$ that allow for the
breaking of rotational invariance of the lattice.

\section{Results}\label{sec:results} 

In what follows, we explore the extent to which the classical scenario is modified by quantum fluctuations 
in the spin-1/2 $XY$ and Heisenberg models. We determine the spin-ordered variational states for 
$0 \leq J_2/J_1 \leq 5$ by performing extensive Monte Carlo simulations for various 
($L \times L \times 2$)-site clusters ($L$ being the number of unit cells along ${\bf a}$ and ${\bf b}$),
with $L=4$, $6$, $8$, $10$, $12$, $14$, $16$, and $18$ ($L=20$ was simulated for a few selected values
of $J_2/J_1$). We also report the energies extrapolated in the thermodynamic limit. The comparison with 
exact results on a small $L=4$ cluster for both the $XY$ and the Heisenberg models is presented in 
the Appendix.

\subsection{Quantum $XY$ model}\label{sec:xymodel}

To identify the presence of possible spirals, one needs to carefully investigate large cluster sizes.
However, even for rather large cluster sizes, only a discrete number of wave vectors are available 
(i.e., ${\bf q}=2\pi/L[n,(2m-n)/\sqrt{3}]$, $n$ and $m$ being integers) and incommensurate spirals 
cannot be captured. Nonetheless, it is possible to reach a quite detailed understanding of the 
evolution of the wave vector describing ordered states.

%%%%%%%%%%%%%%%%%%%%%%%%%%%%%%
\begin{figure}[!t]
\includegraphics[width=1.0\columnwidth]{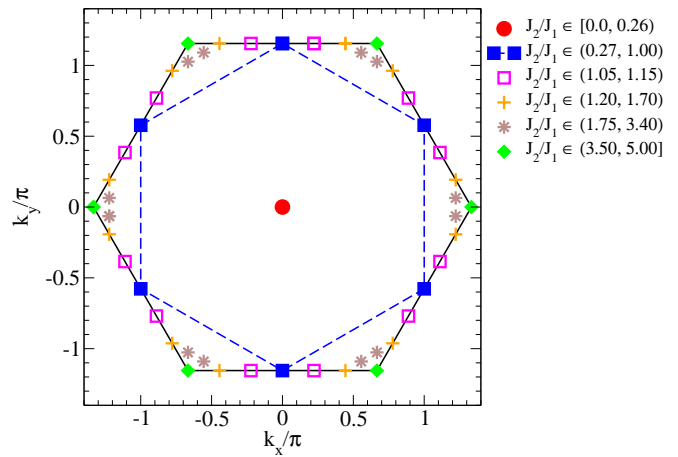} %QXY.eps}
\caption{\label{fig:QXY}
(Color online) The locus of wave vectors for different values of $J_2/J_1$ giving the minimal energy
of the Jastrow variational wave functions. The results are shown for the $XY$ model with $L=18$ 
(i.e., $648$ sites).}
\end{figure}
%%%%%%%%%%%%%%%%%%%%%%%%%%%%%%

\begin{figure}[!b]
\includegraphics[width=1.0\columnwidth]{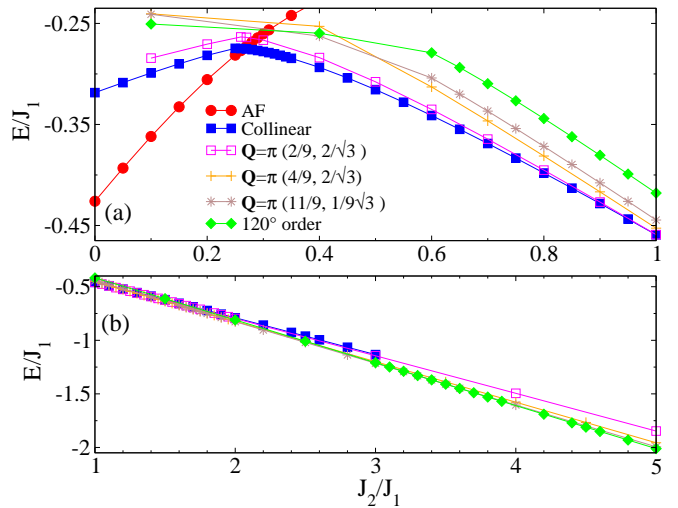} %energyXY.eps}
\caption{\label{fig:energyXY}
(Color online) Variational energies for the $XY$ model for $0 \leq J_2/J_1 \leq 1$ (a) and 
$1 \leq J_2/J_1 \leq 5$ (b) as a function of $J_2/J_1$ for $L=18$. Statistical errors are
smaller than the symbols and solid lines are provided as a guide to the eye.}
\end{figure}

In Fig.~\ref{fig:QXY}, we report the wave vectors ${\bf Q}$ that give the lowest energies of the 
Jastrow states [Eq.~\eqref{eq:Psiansatz}] for different values of $J_2/J_1$. The corresponding energies 
are reported in Fig.~\ref{fig:energyXY} for $0 \leq J_2/J_1 \leq 1$ (a) and 
$1 \leq J_2/J_1 \leq 5$ (b). These results are obtained in clusters with 
$18 \times 18 \times 2$ sites and the typical statistical errors for our energies are of the order 
of $10^{-6}J_1$, i.e., the size of the data points greatly exceeds our error bars.

%%%%%%%%%%%%%%%%%%%%%%%%%%%%%%
%%%%%%%%%%%%%%%%%%%%%%%%%%%%%%

These results show three remarkable effects of quantum fluctuations: (i) a prolongation of the stability
of the N\'eel antiferromagnetic phase with ${\bf Q}=(0,0)$ up to $J_2/J_1 \simeq 0.26$, (ii) the 
extension of the stability of the collinear state from a point at $J_2/J_1 =1/2$ to an entire region 
$0.26 \lesssim J_2/J_1 \lesssim 1$, and (iii) the extension of the stability of the $120^\circ$ state 
from a point at $J_1/J_2=0$ to an entire region with $J_2/J_1 \gtrsim 3.5$. In particular, we find that 
spirals I disappear from the phase diagram and only spirals II occur in the presence of quantum 
fluctuations, between the collinear and the $120^\circ$ states. Moreover, quantum fluctuations determine 
an order-by-disorder lifting of the huge classical degeneracy, the wave vector of spiral phases being 
always along ${\bf M}$-${\bf K'}$ (finite-size effects may favor wave vectors that are close to but not 
exactly along the border zone). Unfortunately, even on the $L=18$ cluster there are only two 
${\bf q}$ points between ${\bf M}$ and ${\bf K'}$ and, therefore, it is extremely difficult to follow 
the evolution of the wave vector for $J_2/J_1 \gtrsim 1$. 

We should add that, as seen in Fig.~\ref{fig:energyXY}, the difference in energy between the competing 
spirals and the collinear and $120^\circ$ states is very small. This suggests the possibility that spirals 
may disappear altogether in the thermodynamic limit, i.e., that quantum corrections favor spirals with 
relatively short periods such as the collinear and $120^\circ$ states. Another possibility is that, as one 
increases the system size, new spirals will appear between the collinear and $120^\circ$ states, pushing
the stability region of the latter state to higher values of $J_2/J_1$.

%%%%%%%%%%%%%%%%%%%%%%%%%%%%%%
\begin{figure}[!t]
\includegraphics[width=1.0\columnwidth]{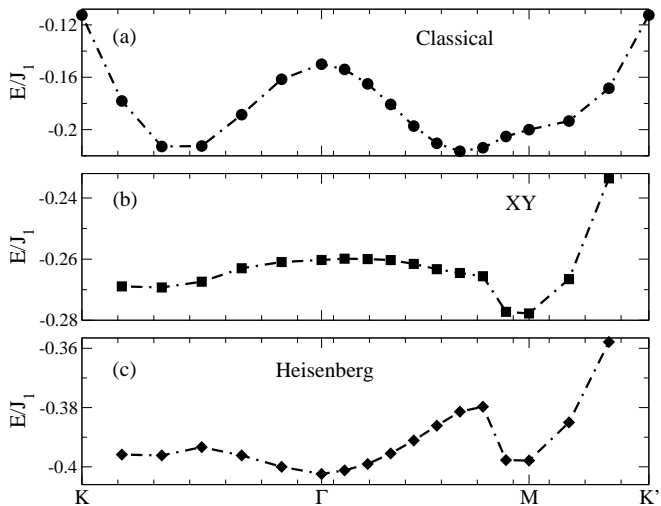} %EqtotJ03.eps}
\caption{\label{fig:eofqj03}
Energy versus wave vector for classical (a), $XY$ (b), and Heisenberg (c) models for $J_2/J_1=0.3$ and 
$L=18$.}  
\end{figure}
%%%%%%%%%%%%%%%%%%%%%%%%%%%%%%

In Fig.~\ref{fig:eofqj03}(b), we report the energy of the $XY$ model as a function of the wave vector 
along the highly symmetric lines in the Brillouin zone for $J_2/J_1=0.3$; the classical results are
reported in Fig.~\ref{fig:eofqj03}(a) for comparison. Figure~\ref{fig:eofqj03}(b) makes apparent that
quantum fluctuations dramatically reduce the dependence of the energy on the wave vectors along these
lines. Furthermore, and more importantly, the minimum in the energy is shifted from spirals I 
(here along ${\bf \Gamma}$-${\bf M}$) to the collinear state with ${\bf Q}={\bf M}$.

%%%%%%%%%%%%%%%%%%%%%%%%%%%%%%
\begin{figure}[!b]
\includegraphics[width=1.0\columnwidth]{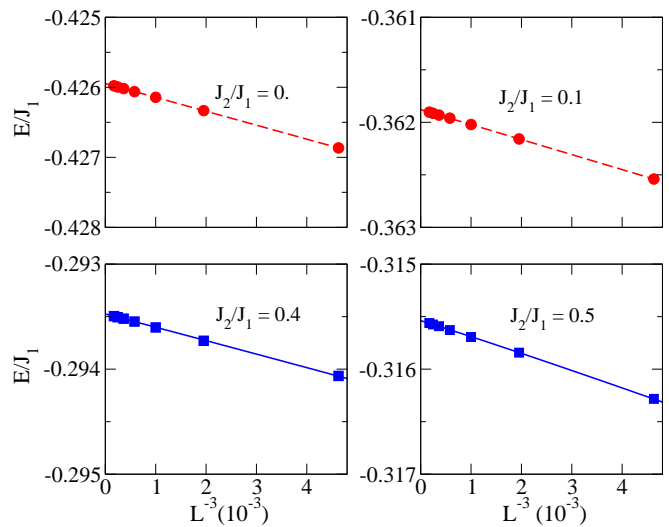} %scalingXY.eps}
\caption{\label{fig:scalingXY}
(Color on-line) Finite-size scaling of the energy of the spin-1/2 $XY$ model for different values of 
$J_2/J_1$ in the N\'eel and collinear phases. Lines depict the results of fits according to 
Eq.~\eqref{eq:scaling}. Red circles (blue squares) indicate the N\'eel (collinear) phase.}
\end{figure}
%%%%%%%%%%%%%%%%%%%%%%%%%%%%%%
\begin{table}[!t]
\centering
\begin{tabular}{|c|l|c|c|c|}
\hline
$J_2/J_1$ & Spin-wave state & $E_\infty$ & $C$ & Classical $E$ \\
\hline
0.0 & N\'eel    & -0.425946(1) & 0.195(1) &  -0.3750000    \\
0.1 & N\'eel    & -0.361879(1) & 0.142(1) &  -0.3000000    \\
0.2 & N\'eel    & -0.305560(2) & 0.092(2) &  -0.2312500    \\
0.3 & Collinear & -0.277761(2) & 0.115(3) &  -0.2166667    \\
0.4 & Collinear & -0.293475(1) & 0.123(2) &  -0.2281250    \\
0.5 & Collinear & -0.315539(1) & 0.138(2) &  -0.2500000    \\
0.6 & Collinear & -0.341104(2) & 0.139(2) &  -0.2770833    \\
0.7 & Collinear & -0.368848(2) & 0.114(3) &  -0.3071428    \\
0.8 & Collinear & -0.398038(2) & 0.075(3) &  -0.3390625    \\
0.9 & Collinear & -0.428247(3) & 0.04(1)  &  -0.3722222    \\
1.0 & Collinear & -0.459214(3) & 0.00(1)  &  -0.4062500    \\
\hline
\end{tabular}
\caption{
Extrapolated energies of the variational Jastrow state in the thermodynamic limit in the spin-1/2
$XY$ model for $0<J_2/J_1<1$. We also report the value of $C$ in Eq.~\eqref{eq:scaling}
and the classical prediction for the energy.}
\label{tab:TabXY}
\end{table}

Having explored the various spin configurations for several cluster sizes, we have also done a 
finite-size scaling analysis of the energy for $0 \le J_2/J_1 \le 1$, where the N\'eel and collinear 
states are found to be the lowest energy states in all sizes considered here. The thermodynamic value
for the energy per site can be obtained from~\cite{Fisher1989,Neuberger1989,Sandvik1998} 
\begin{equation}\label{eq:scaling}
E=E_{\infty} - \frac{C}{L^3} + \frac{D}{L^4},
\end{equation}
where $E_{\infty}$ is the energy in the thermodynamic limit; the fitting parameters $C=\beta c_\text{sw}$
and $D=\alpha c_\text{sw}^2/\rho_s$ provide information about the spin-wave velocity $c_\text{sw}$ and the spin 
stiffness $\rho_s$ ($\alpha$ and $\beta$ are parameters that depend upon the details of the lattice).

Typical examples of the application of this extrapolation procedure are shown in Fig.~\ref{fig:scalingXY}.
For the cluster sizes utilized to determine $E_{\infty}$, Eq.~\eqref{eq:scaling} provides an excellent 
description of the scaling of the data. The nearly linear behavior of the fits shows that the 
contribution of the leading $L^{-3}$ correction to the thermodynamic limit result is dominant for 
the system sizes considered.

The extrapolated results for the energy in the thermodynamic limit, as well as the classical predictions, 
are reported in Table~\ref{tab:TabXY}. A comparison between the two makes apparent that the addition of 
quantum fluctuations dramatically reduces the energy of the ordered states. The values obtained for $C$
[see Eq.~\eqref{eq:scaling}] are also reported in Table~\ref{tab:TabXY}. If one assumes that $\beta$ 
changes moderately with frustration, the evolution of $C$ resembles the behavior of the spin-wave 
velocity $c_\text{sw}$. We find that $C$ has a local minimum around the transition between the N\'eel and the 
collinear phases. Most importantly, our results indicate that the spin-wave velocity remains finite at 
the transition. In addition, $C$ is seen to decrease and vanish as one approaches the transition between 
the collinear state and the spirals, which occurs for $J_2/J_1 \simeq 1$. 

\subsection{Quantum Heisenberg model}\label{sec:heis}

We now study what happens to the ordered phases in the Heisenberg model and show that, in this 
case also, quantum fluctuations strongly modify the classical picture favoring collinear states.

%%%%%%%%%%%%%%%%%%%%%%%%%%%%%%
\begin{figure}[t!]
\includegraphics[width=1.0\columnwidth]{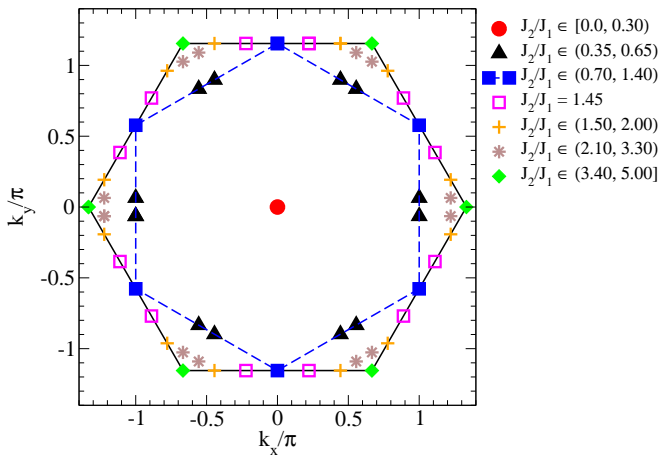} %QHeis.eps}
\caption{\label{fig:QHeis}
(Color online) The same as Fig.~\ref{fig:QXY}, but for the Heisenberg model.}
\end{figure}
%%%%%%%%%%%%%%%%%%%%%%%%%%%%%%

\begin{figure}[!t]
\includegraphics[width=1.0\columnwidth]{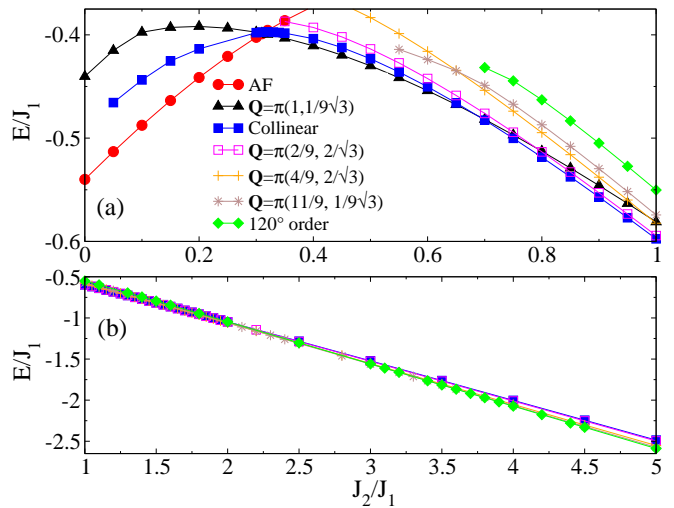} %energyHeis.eps}
\caption{\label{fig:energyHeis}
(Color online) The same as Fig.~\ref{fig:energyXY}, but for the Heisenberg model.}
\end{figure}

In Fig.~\ref{fig:QHeis}, we report the evolution of the wave vector upon increasing frustration on a 
($18 \times 18 \times 2$)-site cluster. The results can be seen to be quite similar to those for the $XY$
model (see Fig.~\ref{fig:QXY}), with an important difference that we highlight in what follows. 
The corresponding energies as a function of $J_2/J_1$ are reported in Fig.~\ref{fig:energyHeis}. 
The results for this large cluster provide strong indications of the trends in the thermodynamic limit. 
First of all, the stability of the N\'eel state persists up to $J_2/J_1 \simeq 0.3$, which is even 
larger than what was found in the $XY$ model. Starting at that point, the best energy is given by the 
spiral with ${\bf Q}=(\pi,0)$: in the $18 \times 18 \times 2$ cluster this wave vector is not present 
and the best energy is found for ${\bf Q}=(\pi,\pi/9\sqrt{3})$, which is the closest point to 
${\bf Q}=(\pi,0)$. For smaller clusters having $(\pi,0)$, we have checked that the spiral with 
${\bf Q}=(\pi,0)$ is indeed the ordered state with the lowest energy. This spiral state is stable up to 
$J_2/J_1 \simeq 0.7$. Therefore, the stability region of this spiral phase is greatly enhanced with 
respect to the classical case, where the $(\pi,0)$ spiral is stable only for $J_2/J_1=0.5$. 
For $J_2/J_1 \gtrsim 0.7$, on the other hand, we find that the collinear state with ${\bf Q}={\bf M}$ 
becomes lower in energy. On further increasing frustration, i.e., for $J_2/J_1 \simeq 1.4$, other spirals 
(with wave vector along ${\bf M}$-${\bf K'}$, similarly to the $XY$ model) become lower in energy. Finally, 
for $J_2/J_1 \gtrsim 3.4$ the $120^\circ$ state is found to be the best variational state.

%%%%%%%%%%%%%%%%%%%%%%%%%%%%%%
%%%%%%%%%%%%%%%%%%%%%%%%%%%%%%

As in the $XY$ model, quantum fluctuations lift the degeneracy between the states with ${\bf Q}=(\pi,0)$ 
and ${\bf Q}=(0,2\pi/\sqrt{3})$. For example, on the $18 \times 18 \times 2$ cluster, the energies of the 
states with ${\bf Q}=(\pi,\pi/9\sqrt{3})$ and ${\bf Q}=(0,2\pi/\sqrt{3})$ are $E=-0.430003(5)$ and 
$E=-0.423142(5)$, respectively, for $J_2/J_1=0.5$; while they are $E=-0.51265(1)$ and $E=-0.51849(1)$, 
respectively, for $J_2/J_1=0.8$. In Fig.~\ref{fig:eofqj03}(c), we report the results for the energy as 
a function of the wave vector for $J_2/J_1=0.3$ and $L=18$. In this case, we can see the appreciable 
differences between the $XY$ and Heisenberg models: in the former, the energy minimum is already at 
${\bf Q}={\bf M}$, while, in the latter, the energy minimum is still at ${\bf Q}={\bf \Gamma}$, and 
local minima can be seen around $(\pi,0)$ and $(0,2\pi/\sqrt{3})$.
  
%%%%%%%%%%%%%%%%%%%%%%%%%%%%%%
\begin{figure}[!t]
\includegraphics[width=1.0\columnwidth]{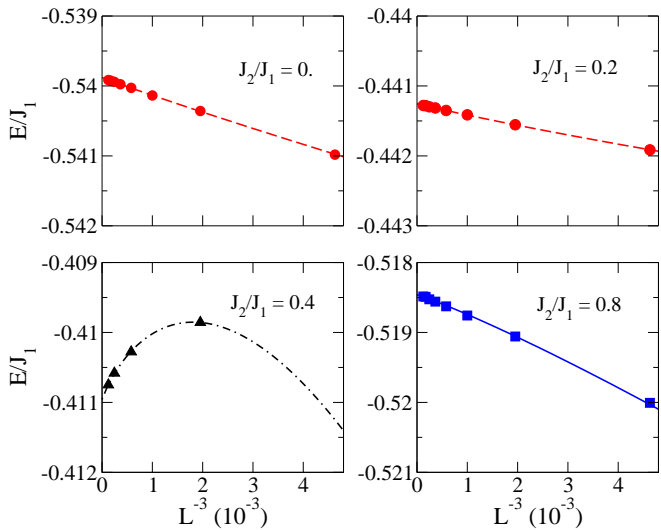} %scalingHeis.eps}
\caption{\label{fig:scalingHeis}
(Color online) The same as Fig.~\ref{fig:scalingXY} but for the Heisenberg model.
Red circles (blue squares) indicate the N\'eel (collinear) phase. The black triangles
indicate the $(\pi,0)$ spiral.}
\end{figure}
%%%%%%%%%%%%%%%%%%%%%%%%%%%%%%

By comparing the results for the $XY$ (see Fig.~\ref{fig:energyXY}) and Heisenberg (see 
Fig.~\ref{fig:energyHeis}) models one can see that the main qualitative difference between them is 
the existence of a $(\pi,0)$ spiral as the lowest energy state in the latter. Hence, while quantum 
fluctuations lift the degeneracy between the $(\pi,0)$ spiral and the collinear state in both models, 
only in the Heisenberg model do these spirals appear as the ground state in our variational approach based 
upon Jastrow wave functions.

In Fig.~\ref{fig:scalingHeis}, we show typical examples of the extrapolations done for the Heisenberg 
model in order to obtain the energy in the thermodynamic limit, where we have used the scaling in 
Eq.~\eqref{eq:scaling}. As for the $XY$ model (see Fig.~\ref{fig:scalingXY}), we have found that 
Eq.~\eqref{eq:scaling} provides an excellent description of the data for the N\'eel and collinear states.
The nearly linear behavior of the fits makes apparent that the contribution of the leading $L^{-3}$ 
correction to the thermodynamic result is dominant for the system sizes considered. In contrast, the 
finite-size trend for the $(\pi,0)$ spiral shows an anomalous behavior, with a positive slope (i.e., 
a negative coefficient $C$). Moreover, in this case, a substantial contribution from the sub-leading 
corrections $L^{-4}$ is present, i.e., our results for this state (which include the cluster 
with $L=20$) suffer from strong finite-size effects. Hence, larger clusters and/or Jastrow factors with 
higher-order correlations are needed to clarify the fate of the $(\pi,0)$ spiral in the thermodynamic limit.

Results for the extrapolated energies and corresponding orderings for the Heisenberg model in the 
thermodynamic limit are shown in Table~\ref{tab:TabHeis}, where they can be compared to the classical
predictions. As for the $XY$ case, one can see that the addition of quantum fluctuations reduces the energy
dramatically. In Table~\ref{tab:TabHeis}, we also report the values of the fitting parameter $C$ whenever
it is positive. As discussed for the $XY$ model, one can obtain information on the behavior of the 
spin-wave velocity $c_\text{sw}$ from the coefficient $C$ of the fitting procedure of Eq.~\eqref{eq:scaling}. 
Here, $C$ decreases upon increasing frustration but does not vanish at the transition from the 
antiferromagnetic state to the $(\pi,0)$-spiral. This behavior is similar to that observed in the $XY$ 
model for the transition between the antiferromagnetic state and the collinear one.

\begin{table}[!t]
\centering
\begin{tabular}{|c|l|c|c|c|}
\hline
$J_2/J_1$ & Spin-wave state & $E_\infty$ & $C$ & Classical $E$ \\
\hline
0.0 & N\'eel           & -0.539880(2) & 0.280(3) &  -0.3750000 \\
0.1 & N\'eel           & -0.487544(2) & 0.230(3) &  -0.3000000 \\
0.2 & N\'eel           & -0.441254(2) & 0.191(4) &  -0.2312500 \\
0.3 & N\'eel           & -0.402398(3) & 0.127(4) &  -0.2166667 \\
0.4 & $(\pi,0)$ spiral & -0.410958(5) &          &  -0.2281250 \\
0.5 & $(\pi,0)$ spiral & -0.43051(1)  &          &  -0.2500000 \\
0.6 & $(\pi,0)$ spiral & -0.45476(1)  &          &  -0.2770833 \\
0.7 & $(\pi,0)$ spiral & -0.48261(1)  &          &  -0.3071428 \\
0.8 & Collinear        & -0.518460(3) & 0.198(6) &  -0.3390625 \\
0.9 & Collinear        & -0.556993(4) & 0.153(8) &  -0.3722222 \\
1.0 & Collinear        & -0.597471(4) & 0.084(9) &  -0.4062500 \\
\hline
\end{tabular}
\caption{
Extrapolated energies of the variational Jastrow state in the thermodynamic limit in the spin-1/2
Heisenberg model for $0<J_2/J_1<1$. We also report the value of $C$ in Eq.~\eqref{eq:scaling},
and the classical prediction for the energy.}
\label{tab:TabHeis}
\end{table}

\section{Conclusions}\label{sec:conc} 

We have explored the stability of classically ordered states within the phase diagram of the spin-1/2 $XY$ 
and Heisenberg models on the honeycomb lattice, using long-range Jastrow wave functions and Monte Carlo 
simulations. 

For the $XY$ model, in the context of our variational calculations, we find that quantum fluctuations 
extend the stability of the antiferromagnetic state up to $J_2/J_1\simeq 0.26$ (to be compared to 
$J_2/J_1=1/6$ in the classical case), of the collinear state for $0.26\lesssim J_2/J_1\lesssim 1$ (to be 
compared to $J_2/J_1=0.5$ in the classical case), and of the $120^\circ$ phase for $J_2/J_1\gtrsim 3.5$ 
(to be compared to $J_2/J_1=\infty$ in the classical case). Quantum fluctuations are found to suppress 
spirals I, while spirals II still occur for $1\lesssim J_2/J_1\lesssim 3.5$, in between the collinear and 
$120^\circ$ states. Since the difference in energy between spirals II and the collinear or $120^\circ$ states 
is very small in the region where the former are the lowest-energy states in the finite clusters studied, an 
open question is whether those spirals remain stable in the thermodynamic limit, or whether they disappear 
and a direct transition occurs between collinear and $120^\circ$ states. 

A comparison between the results of our variational calculations for spirals in the spin-1/2 $XY$ model 
to those of exact diagonalization (see the Appendix), partonic wave function 
studies~\cite{Carrasquilla2013}, and density-matrix renormalization group calculations~\cite{Zhu2013b} 
makes apparent that in-plane magnetically ordered states are not expected to appear as ground states in the 
maximally frustrated region $0.2\lesssim J_2/J_1\lesssim 0.36$. On the contrary, magnetically ordered states 
described by Eq.~\eqref{eq:Psiansatz} are expected to correctly describe the ground-state properties for
both $J_2/J_1 \lesssim 0.2$ and probably $J_2/J_1 \gtrsim 0.36$ (for the latter case, more work is needed
to understand the precise value of $J_2$ at which magnetic long-range order occurs).

For the Heisenberg model, we find that quantum fluctuations extend the stability of the antiferromagnetic 
state up to $J_2/J_1 \simeq 0.3$ (to be compared to $J_2/J_1=1/6$ in the classical case), of the 
$(\pi,0)$ spiral state for $0.3 \lesssim J_2/J_1 \lesssim 0.7$ (to be compared to $J_2/J_1=0.5$ in the 
classical case), of the collinear state for $0.7 \lesssim J_2/J_1 \lesssim 1.4$ (to be compared to 
$J_2/J_1=0.5$ in the classical case), and of the $120^\circ$ phase for $J_2/J_1\gtrsim 3.4$ (to be compared 
to $J_2/J_1=\infty$ in the classical case). Spirals II appear for $1.4\lesssim J_2/J_1\lesssim 3.4$, 
 between the collinear and $120^\circ$ states. As in the $XY$ case, since the difference in energy between 
spirals II and the collinear and $120^\circ$ states is very small, their stability in the thermodynamic 
limit remains uncertain. 

We note that, for our finite-size calculations of the spin-1/2 $XY$ and Heisenberg models, quantum 
fluctuations lift the macroscopic classical degeneracy of spirals II favoring wave vectors along the 
border zone ${\bf M}$-${\bf K'}$, in agreement with $O(1/S)$ spin-wave calculations~\cite{Mulder2010}. 
In addition, they lift the degeneracy between the $(\pi,0)$ spiral state and the collinear state. As a 
result, the former state disappears from the phase diagram of the $XY$ model while its stability
is enhanced in the Heisenberg one. However, we found that our results for the energy of the $(\pi,0)$-spiral 
state in the Heisenberg model exhibit strong finite-size effects, which suggests that they need to be 
reconsidered with calculations on larger clusters and/or including Jastrow factors with higher-order 
correlations.

%%%%%%%%%%%%%%%%%%%%%%%%%%%%%%
\begin{figure}[t!]
\includegraphics[width=1.0\columnwidth]{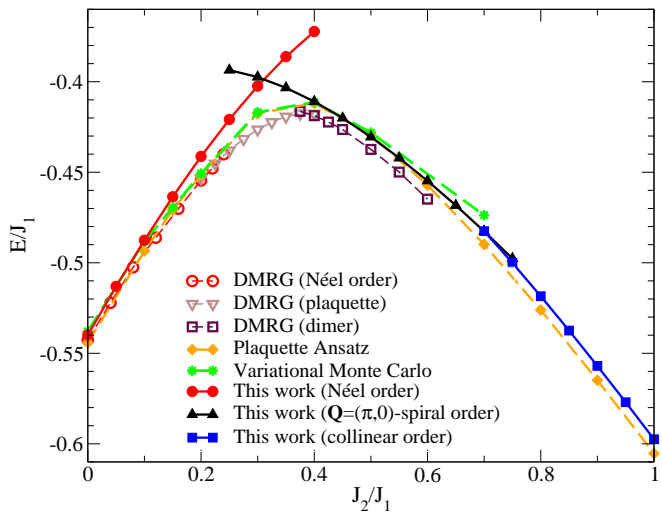} %comparison.eps}
\caption{\label{fig:comparison}
(Color online) Comparison between our variational approach for magnetically ordered states and other
numerical approaches that have been considered recently.}
\end{figure}
%%%%%%%%%%%%%%%%%%%%%%%%%%%%%%

Finally, we would like to briefly discuss the relation of our variational calculations of magnetically
ordered phases with previous numerical calculations on the spin-1/2 Heisenberg model. 
In Fig.~\ref{fig:comparison}, we report the energy of our best states together with recent density-matrix
renormalization group (DMRG) calculations~\cite{Ganesh2013}, variational approaches based upon Jastrow
and projected fermionic states~\cite{Clark2011}, or plaquette {\it Ans\"{a}tze}~\cite{Mezzacapo2012}. 
(A comparison with exact diagonalization results for a 32-site cluster is presented in the
Appendix). In general, ordered states provide accurate approximations for the exact 
ground state for small and large values of $J_2/J_1$. Especially for $J_2/J_1 \lesssim 0.2$, our energies 
are competitive with the DMRG ones, indicating that the N\'eel ordered phase occurs in that regime. 
The DMRG result that the N\'eel phase is obtained beyond its classical stability region is fully
compatible with our present results, showing that (in contrast to other results for frustrated lattices such as,
for example, the $J_1$-$J_2$ model on the square lattice) quantum fluctuations can reinforce collinear
magnetic order.

On the contrary, our spin-wave states have a rather poor accuracy in the highly frustrated regime 
$0.2 \lesssim J_2/J_1 \lesssim 0.4$, where magnetically disordered phases with plaquette and dimer 
order should occur~\cite{Ganesh2013,Zhu2013a,Gong2013}. The $(\pi,0)$ spiral state  may disappear 
altogether, being replaced by disordered plaquette and dimer phases~\cite{Ganesh2013,Zhu2013a,Gong2013}. 
For larger values of $J_2/J_1$, our ordered states become again competitive with other those from methods, 
indicating that magnetically ordered phases are present in that part of the phase diagram. In particular, the
collinear phase with ${\bf Q}=(0,2\pi/\sqrt{3})$ (and related ones) may be relevant for 
$J_2/J_1 \gtrsim 0.8$, where DMRG results~\cite{Ganesh2013} showed clear evidence of a rotational-symmetry 
breaking and possibly a vanishing spin gap.

\section{Acknowledgments}
This work was supported by the National Science Foundation under Grant No.~OCI-0904597 (A.D.C. and M.R.), 
the Office of Naval Research (J.C. and M.R.), PRIN 2010-11 (F.B.), and the U.S. ARO (V.G.). J.C. acknowledges 
support from the John Templeton Foundation. Research at Perimeter Institute is supported by the Government 
of Canada through Industry Canada and by the Province of Ontario through the Ministry of Research \& Innovation. 
We thank A. Parola for providing us with the exact results on the $4 \times 4 \times 2$ cluster
shown in the Appendix; R. Ganesh for his DMRG results shown in Fig.~\ref{fig:comparison}; and C. N. Varney,
K. Sun, and S. Nishimoto for stimulating discussions.

\appendix

\section{Comparison with exact results on a small cluster}\label{sec:appendix}

In order to test the accuracy of our spin wave states, here we present a direct 
comparison between variational and exact energies on a small ($4 \times 4 \times 2$)-site cluster for
$0 \le J_2/J_1 \le 1$. The results for the energy are reported in Tables~\ref{tab:TabXY4x4} 
and~\ref{tab:TabHeis4x4} for the $XY$ and Heisenberg models, respectively. In such a cluster, only a few 
spirals can be accommodated (there are only four independent momenta available). We mention that the exact 
ground state is always in the ${\bf Q}=(0,0)$ subspace, for both the $XY$ and the Heisenberg models.

\begin{table}[!t]
\centering
\begin{tabular}{|c|l|c|c|c|}
\hline     
$J_2/J_1$ & Spin-wave state & $E_{\rm Jastrow}$ & $E_{\rm exact}$ & Accuracy \\
\hline  
0.0 & N\'eel               & -0.429079(1) & -0.4294059  & 0.001 \\
0.1 & N\'eel               & -0.364117(2) & -0.3654374  & 0.004 \\
0.2 & N\'eel               & -0.306949(4) & -0.3149448  & 0.025 \\
0.3 & Collinear            & -0.279516(4) & -0.2952758  & 0.053 \\
0.4 & Collinear            & -0.295509(3) & -0.3016051  & 0.020 \\
0.5 & Collinear            & -0.318238(3) & -0.3237236  & 0.017 \\
0.6 & Collinear            & -0.344366(3) & -0.3502100  & 0.017 \\
0.7 & Collinear            & -0.372495(4) & -0.3789541  & 0.017 \\
0.8 & Collinear            & -0.401916(4) & -0.4090846  & 0.017 \\
0.9 & Collinear            & -0.432229(4) & -0.4401615  & 0.018 \\
1.0 & Collinear            & -0.463183(5) & -0.4719334  & 0.019 \\
\hline   
\end{tabular}    
\caption{
Comparison between the exact ground-state energies and the energies of the best Jastrow wave 
function on the $4\times 4 \times 2$ cluster for the $XY$ model.}
\label{tab:TabXY4x4}
\end{table}
On the $XY$ model, the agreement between the variational and the exact ground-state energies is 
excellent for the unfrustrated case (where it is about $0.1\%$). Then the accuracy deteriorates on
increasing $J_2/J_1$. The worst results are obtained in the highly frustrated regime 
$0.2 \le J_2/J_1 \le 0.4$, where a magnetically disordered phase is expected to occur in the
thermodynamic limit. Then the accuracy is remarkably good also for $J_2/J_1 \gtrsim 0.4$, even though
the variational state has a finite wave vector. 
\begin{table}[h!]
\centering
\begin{tabular}{|c|l|c|c|c|}
\hline
$J_2/J_1$ & Spin-wave state & $E_{\rm Jastrow}$ & $E_{\rm exact}$ & Accuracy \\
\hline
0.0 & N\'eel           & -0.54528(1) & -0.5516867  & 0.012 \\
0.1 & N\'eel           & -0.49168(1) & -0.4998728  & 0.016 \\
0.2 & N\'eel           & -0.44428(1) & -0.4567175  & 0.027 \\
0.3 & Collinear        & -0.40600(1) & -0.4275835  & 0.050 \\
0.4 & $(\pi,0)$ spiral & -0.41217(1) & -0.4203210  & 0.019 \\
0.5 & $(\pi,0)$ spiral & -0.43401(1) & -0.4424316  & 0.019 \\
0.6 & $(\pi,0)$ spiral & -0.46160(1) & -0.4738271  & 0.026 \\
0.7 & $(\pi,0)$ spiral & -0.49363(1) & -0.5103811  & 0.033 \\
0.8 & $(\pi,0)$ spiral & -0.52908(1) & -0.5500986  & 0.038 \\
0.9 & $(\pi,0)$ spiral & -0.56719(1) & -0.5919934  & 0.042 \\
1.0 & $(\pi,0)$ spiral & -0.60742(1) & -0.6355307  & 0.044 \\
\hline
\end{tabular}
\caption{
Comparison between the exact ground-state energies and the energies of the best Jastrow wave
function on the $4\times 4 \times 2$ cluster for the Heisenberg model.}
\label{tab:TabHeis4x4}
\end{table}
On the Heisenberg model, the trend is similar to the previous one, even though the actual accuracy is 
always lower than in the $XY$ model. The best variational energies are obtained in the unfrustrated
regime (about $1.2\%$). Again, the worst accuracy appears for $J_2/J_1=0.3$, where a non-magnetic
phase is expected to occur in the thermodynamic limit. Then a substantial improvement is obtained
for $J_2/J_1=0.4$, but from there, the accuracy monotonically deteriorates on increasing the frustrating 
ratio, still remaining below $5\%$ up to $J_2/J_1=1$. Note that, for larger clusters, the N\'eel state 
has a lower energy than the collinear state for $J_2/J_1=0.3$, while the collinear 
state has a lower energy than the $(\pi,0)$-spiral for $J_2/J_1\gtrsim 0.7$.

\end{document}